\documentclass[sigconf]{acmart}

\usepackage{multirow} 
\begin{document}

\title{Transferring Studies Across Embodiments: A Case Study in Confusion Detection}


\author{Na Li}
\affiliation{%
  \institution{School of Computer Science\\
  Technological University Dublin  \\ }
  \city{Dublin}
  \country{Ireland}}
\email{na.li@tudublin.ie}

\author{Robert Ross}
\affiliation{%
  \institution{ School of Computer Science\\
  Technological University Dublin  \\ }
  \city{Dublin}
  \country{Ireland}}
\email{robert.ross@tudublin.ie}

\begin{abstract}
Human-robot studies are expensive to conduct and difficult to control, and as such researchers sometimes turn to human-avatar interaction in the hope of faster and cheaper data collection that can be transferred to the robot domain. In terms of our work, we are particularly interested in the challenge of detecting and modelling user confusion in interaction, and as part of this research programme, we conducted situated dialogue studies to investigate users' reactions in confusing scenarios that we give in both physical and virtual environments. In this paper, we present a combined review of these studies and the results that we observed across these two embodiments. For the physical embodiment, we used a Pepper Robot, while for the virtual modality, we used a 3D avatar. Our study shows that despite attitudinal differences and technical control limitations, there were a number of similarities detected in user behaviour and self-reporting results across embodiment options. This work suggests that, while avatar interaction is no true substitute for robot interaction studies, sufficient care in study design may allow well executed human-avatar studies to supplement more challenging human-robot studies.
\end{abstract}

\ccsdesc[100]{Human-Computer Interaction~Avatar}
\ccsdesc[100]{Human-Robot Interaction~Social Robot}
\keywords{human-computer interaction, human-robot interaction, confusion detection, wizard-of-oz, pepper robot, avatar, user engagement}

\maketitle

\section{Introduction}

Due to the relative ubiquity of computer mediated communication across different application domains \textit{e.g.} online learning system \cite{engagementHCI2019}, healthcare assistants \cite{healcarehumanrobot2020}, virtual reality (VR) games \cite{VRgameAvatar2019}, and social VR \cite{ACM2021olderpeople}), the expectations for multimodal interactive systems have grown in diversity and sophistication in the last decade. This is true for virtual online agents, but also extends to expectations for interaction with physical, or more precisely social, robots \cite{engagementHCI2019,mulitmodalSocialtecholiques2021}. A key trend in the development of communicative systems has been an assumption of multi-modality, \textit{i.e.}, that our artificial interlocutors should have access to multiple modalities. However, the research community is well aware that the multimodal communicative skills of even state-of-the-art systems are still very limited. 

Whether our interaction partner is a social robot, a 3D avatar, or even just a chat window, our interaction partners are assumed to share similar conversational skills and abilities across these embodiments. This has benefits in terms of acclimatisation of technology across interaction partner types, but can also lead to frustration and disappointment when such alignment is not present in practise. This though is not just true in terms of users expecting systems to behave in uniform ways across device type, but may also be present in the expectations that systems -- and their designers -- make in terms of the behaviour and reaction of users to systems across different embodiment types. 

This potential for mismatched expectations is in some case exacerbated by the needs of researchers and industrial developers. Collecting real world data in HRI studies for the investigation of particular phenomena is extremely challenging. Experimental hardware systems suffer malfunctions, recruiting participants or users is challenging and often expensive, and even finding appropriate real world spaces to perform tasks can be difficult. These problems were significantly heightened during the COVID-19 pandemic when it became in many ways unfeasible to perform human-robot interaction studies. For reasons such as these numerous researchers have over the last four decades frequently turned to human-computer interaction studies, and particularly the use of avatars and chat systems, to conduct studies in the hope of bootstrapping studies of HRI. Invariably these efforts have been based on an assumption that such data is as ecologically valid in one embodiment type as in another so long as the same basic interaction modalities are being used, \textit{e.g.}, speech and cameras. While this assumption may have been true at one point, due to the relative novelty of all interactive systems interfaces, the ubiquitous nature of avatars and basic conversational systems in contrast with everyday social robotics, has arguably laid waste to this assumption. 

In light of the above argument, in this paper we present a contrastive analysis of the reactions of users across the physical robot and avatar embodiment types in order to investigate if it remains feasible to leverage human-avatar data for human-robot interaction when the focus of these studies is on communication rather than for example physical cooperation. This study has been executed with respect to our core focus of interest which is the modelling and mitigation of user confusion in social robotics tasks. For the avatar virtual embodiment option, we analyse data from a previously presented 3D avatar study. For the physical robot we present for the first time our study using a humanoid Pepper robot to execute a similar multimodal situated conversational study. Our comparison across these two studies aims to answer the following two questions: \textbf{(1)} To what extent can we rely on human-avatar interaction (HAI) studies as a substitute for human-robot interaction (HRI) data collection efforts? \textbf{(2)} Does the choice of embodiment type (HAI vs. HRI) have a significant effect on users perceptions and mental state with respect to our central challenge of confusion detection? 

\section{Related Work}
\label{sec:relatedwork}

Before presenting details of the studies, we begin with a review of related work with a brief review of some notable studies on HCI and HRI, differences seen between embodiment types, and some work related to our case study on confusion detection.  

\subsection{Embodied Interaction}

While Human-Computer Interaction (HCI) covers a vast number of physical system types as well as different goals of interaction, we are particularly interested in situated interaction where a user communicates with an embodied agent which is typically a physical robot, but can also be embodied virtually in our study. In recent years, avatars have begun to become quite prominent as a mechanism in intelligent virtual environments \cite{trustAvatarorrobot2016}. Compared to other means of interaction, the avatar is presumed to be a more natural communication mechanism which can evoke strong agent-as-partner style interactions through their use of human-like facial features and expression \cite{avatarDialogue2017}, vivid body language, and even specific personalities. Moreover, the avatar has remarkable benefits over a speech-to-text-only interaction \cite{avatarDialogue2017}. 

In the case of developing human-robot interaction, the challenge of combining multiple modalities in addition to the purely spoken channel has been long recognised. A social robot should have the ability to recognise affect, emotion, or engagement by observing the behaviour of the interlocutor, such as tracking the gaze, the head pose, facial expressions, gestures, or biological behaviours (\textit{e.g.} heartbeat, electroencephalography activity (EEG) and body temperature) \cite{eyegazing2017,affectrecognition2014,IEMOCAPBusso2008,emotionexpressioniHRI2019,reviewVerbalNonHRI2015}. Indeed, increasing the affordance of human-like conversation in HRI has in particular been the subject of increased research over the last two decades \cite{hribook2020,socialinteraction_HRI}. 

Multiple studies have commented on the relative properties of communication with physical robots versus other types of agents. Generally, it has been observed that people have more interactions with physical robots than with virtual agents or telecommunication agents in a number of different application areas \cite{socialinteractiveHRI2006,statisfiedHRI2006}. Meanwhile, \citet{affitivediffhrihci2006} revealed that customers or users can be affected to varying degrees in their overall user experience, due to perception of different levels of social presence across HCI and HRI. In \citet{sameenginetworobots2020}'s study of social presence, they approached HCI and HRI experiments with the same conversational engine but with keyboard and monitor used for the HCI studies, and a Nao robot used for the HRI studies. In post-questionnaires of participants, in particular, the ``UTAUT'' (Unified Theory of Acceptance and Use of Technology Questionnaire \cite{utautquestionnaire2010}), it was shown that HRI trended more strongly with the measures of animacy and likeability than with the HCI, while on the measure of usefulness and trust, the HCI experience was rated higher than in the HRI case. The authors believed that the HCI performance is better than the HRI performance in specific tasks or domains, but that the HRI performance was better than the HCI for the exploratory and open-ended conversation domains. Also of note, in a cooperative blockstick HRI / HCI task, participants were found to be more engaged and better enjoyed playing with a physically embodied robot compared to playing with a virtual embodied animated avatar as the physical robot was informative and creditable \cite{effectrobot2004}. On the contrary, \citet{effectrobot2004} also found that, such as verbal and role-playing tasks, there is no significant difference in user attitudes between users who interact directly with a robot and who play with the robot video-displayed remotely in different room. 

\subsection{Confusion Detection}

For HRI to become more natural, it is essential that systems be tuned to adapt to the user's mental state during task progression. Confusion is a mental state that is characterised by bidirectional emotion, which means it can be a positive engagement in a conversation prior to a learning event, but it can also be correlated with negative states (such as boredom or disengagement) in an interaction \cite{DMelloConfusionlearning2014,li2021detecting}. This more negative view on confusion looks on it as an epistemic emotion \cite{Lodge2018} which is associated with blockages or impasses in the learning process when trying to learn something new or trying to clarify problems. There have been multiple formal definitions of confusion within the HCI and psychology literature \cite{Arguel2015,LEHMAN2012184,Silvia2010ConfusionAI}, and within the HCI literature there have been multiple studies into confusion detection \cite{eegconfusionlevel,Grafsgaard2011,Zhou2019}. However, the majority of study in this area has concerned online learning systems, and little work has focused on general engagement or task-orientated interaction in HRI.

Building on a newly defined framework for confusion study, \citet{li2021detecting} conducted a study specifically to investigate the manifestation and detection of confusion states in the context of HAI. For this HAI confusion detection study, a web application framework was developed; this framework includes a real-time web application and an avatar application based on that presented by \citet{Sloan2020Emotional}. Data were gathered from 19 participants (8 males, 11 females) from six countries. The study used a three-task design in which users were presented with confusing or non-confusing variants of particular tasks and were assessed using video analysis, speech analysis, and assessment of feedback forms. 

\section{Study and Data}

Studies discussed in Section \ref{sec:relatedwork} show that physical robots may have greater social attribution and enjoyment than artificial virtual avatars \cite{Bainbridge2011}, but that there is little systematic investigation of differences in performing experiments across different embodiments. Given this and our particular focus on the challenge of confusion detection in interaction, we have built on our earlier study by conducting a follow-up HRI study with similar study goals and methods to investigate the cross embodiment differences observed. Here, we present an overview of the methods used for conducting the HRI study, and in the following section a systematic companion of our study results across the two modalities.  

\section{Study Design}

In our earlier HAI study, we made use of a Wizard-of-Oz (WoZ) \cite{Riek2012WizardOO} mechanism to conduct user studies and collect data, which we later subjected to a series of manual, semiautomatic, and fully automatic methods to analyse the behaviour of participants \cite{li2021detecting}. As with the HAI study, for our HRI study, we used a semi-spontaneous one-to-one WoZ conversation between the Pepper robot and a participant. Each participant took less than 15 minutes in total, with 5 minutes for the task-centred part of the conversation. At the beginning of each experiment, participants received consistent instructions and consent forms as with the HAI study. The study itself consisted of three tasks that the participant had to try to complete -- these could be presented in confusing or non-confusing configurations. After completing the tasks, these participants were asked to complete a post-questionnaire. At the end of the experiment, each participant was invited for a 3-minute interview with the researcher.

Regarding the dialogue design, two conditions are defined for the corresponding appropriate stimuli. Condition A is related to confusion of the stimuli, while condition B is that the stimuli were designed so that participants could complete straightforward tasks. We designed a set of conversations with three tasks for the two studies: task 1 was a logical problem; task 2 was a word problem; while task 3 was a maths problem. For each task, we designed two conversations with the two conditions with four patterns \cite{li2021detecting} (complex-simple information, contradictory-consistency information, insufficient-sufficient information, and feedback). 

\begin{figure}
	\includegraphics[width=0.5\textwidth]{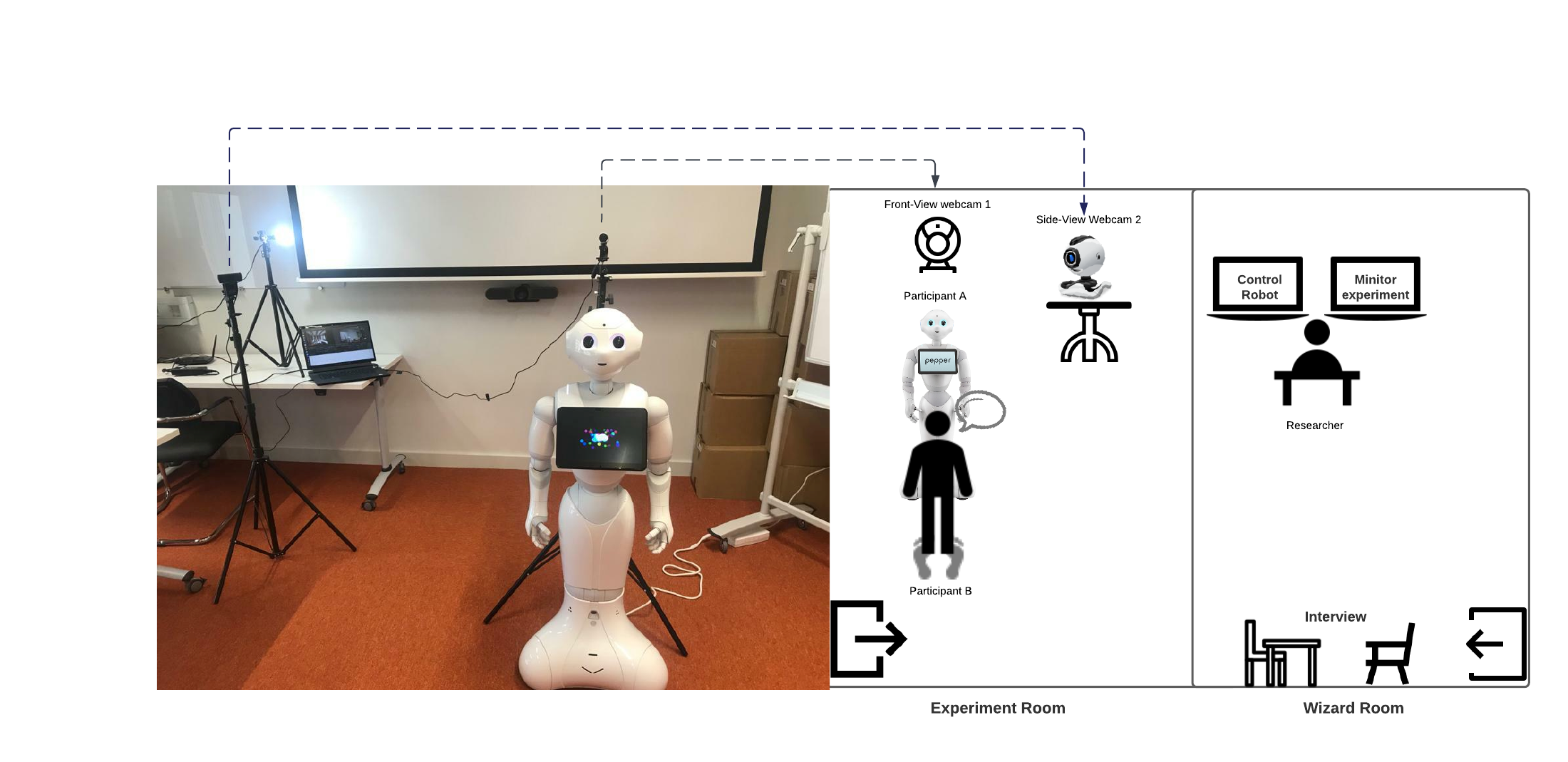}
	\caption{\label{fig:hriexperimentroom} HRI Laboratory Setup}
\end{figure}

For the HRI experimental design, we prepared two experiment rooms (see Figure \ref{fig:hriexperimentroom}): the experiment room and the wizard room. Only the Pepper robot and a participant are in the experiment room. There are two HD webcams setup:
Webcam 1 was used to record high quality views of the participant's facial expression; while Webcam 2 is further back, but again behind the Pepper robot to record participants' body language. The left picture shows the actual laboratory setup of the experiment. We also provided additional lighting for the participants to ensure high-quality video was collected.
In the wizard room, the researcher controlled the robot using a WoZ4U platform \cite{WoZ4U2021}, which is an open source web application that provides a graphical user interface (GUI) for the wizard to control the Pepper robot. We integrated our experiment scripts and the Pepper robot behaviours on the WoZ4U. For example, participants can scan a feedback QR code on the Pepper tablet; designed conversation scripts with the animated speech of the Pepper robot (\textit{e.g.}, happy, embarrassed, wave, \textit{etc.}).

\begin{table*} 
    \caption{\label{tab:haihriemtion_7}The Result of emotion estimation grouped by two conditions in HAI and HRI}    
    \begin{tabular}{l|l|l|l|l|l|l|l|l|l}
    \hline \textbf{Modality} &\textbf{Condition} & \textbf{Anger} & \textbf{Disgust}  & \textbf{Fear} & \textbf{Sadness} & \textbf{Happiness}  & \textbf{Surprise}& \textbf {Neutral} & \textbf{Overall}\\ \hline
    \multirow{2}{*}{HAI} & A  &  262 & 282 & 136 & 677 & 702 & 65 & 1799 & 3923\\
     & B  &  77 & 165 & 57 & 480 & 858 & 95 & 1502 & 3234 \\
      \hline
      \multirow{2}{*}{HRI} & A  &  40 & 62 & 1511 & 59 & 91 & 67 & 92 & 1922\\
      & B  &  19 & 46 & 1503 & 57 & 151 & 48 & 102 & 1926 \\
  \hline
\end{tabular}
\end{table*}

\section{Data Acquisition and Analysis}

In the HRI study, there were 29 participants (16 males, 12 females, 1 not stated) with different educational and ethnic backgrounds. The total HRI dataset includes audio data that were recorded from the Pepper's microphone, two facial videos, one from the Pepper robot's camera, and another one Webcam 1, and another posture video. The current analysis mirrors the approach taken with the earlier HAI study. Specifically, frame data were extracted only from facial video data. The post-study questionnaire included 10 questions using a 5-level Likert scale. Three questions were about the three tasks, including confusion scores for the two conditions. The other three questions were specific positive feedback and the other three questions were toward negative feedback. The last question is about whether the participants would abandon interactions with the robot. Generally, we found that there were minimal data collection issues in the HRI study compared to our earlier HAI study, since the researcher controls all variables in the HRI setup. 

We applied several feature analysis algorithms to facial frame data and survey feedback to investigate the different behaviour of participants between Condition A and Condition B, and subsequently analysed the results of these variables between the current HRI study and the previous HAI investigation. From the facial frame data, we extracted indicators of emotion, eye gaze, and head pose. For emotion detection, we used a visual emotion detection algorithm \cite{facialemotion2021} that was used to estimate 7 target classes (Neutral, Happy, Sad, Surprise, Fear Anger, and Disgust). For eye gaze estimation, a state-of-the-art eye gaze estimation model was applied. This algorithm is trained on more than one million high-resolution images with different gazes in extreme head poses \cite{Zhang2020ETHXGaze}. For head-pose estimation, we applied the model presented by \citet{headpose2017}; This work uses CNNs, dropout and adaptive gradient methods trained on three novel datesets \cite{li2021detecting}.

\section{Results}

To usefully compare the results of the HAI and HRI, we repeat relevant portions of the results of the HAI study here in addition to our new HRI results. In the case of emotions detected, in the HAI experiment, the predicted results in Condition A corresponding to the four classes of negative emotions (anger, disgust, fear, and sadness) are stronger than in the case for these classes in Condition B. In contrast, the number of predicted results for the two positive emotions (happiness and surprise) in Condition A is less than in Condition B. In the HRI experiment, we can see that the results of the five main predicted emotions are slightly similar to the HAI experiments, except for the two special surprise and neutral emotions (see Table \ref{tab:haihriemtion_7}).

Considering gaze estimation, through an independent-samples t-test of results from the HAI and HRI studies, we found that there is a significant difference in the sum of absolute values of pitch and yaw across the two conditions for HAI and HRI experiments, respectively. In the case of head pose, the independent-samples t-test result shows there is a significant relationship between the sum of absolute values of three angles (pitch, yaw, and roll) and two conditions for the HAI study; while there is no significant difference between the sum of absolute values of these three angles and two conditions for the HRI study.

Regarding subjective self-reporting scores on 48 participants, we analysed the self-estimated confusion scores of each participant, and the user attitude towards the embodiment option (avatar vs. social robot). For both HAI and HRI options, a significant difference was found between confusing and non-confusing tasks only in the case of Task 3 (maths problems). There was no significant correlation in the confusion scores for task 1 (logic problems) and task 2 (word problems) for conditions A and B in the HAI or HRI studies. However, what is more interesting is in the analysis of the user experience questions: this included an examination of whether there is a significant difference between the average scores of negative feedback of the user's experiences in the two studies; an examination of whether there is a significant relationship between the average scores of positive feedback of the users' experiences in the two studies; and finally, an examination of whether there is a statistically significant relationship between the score of participants wanting to abandon these conversations and these two studies.

We found that there is a significant difference between the average negative feedback scores in the two studies ($M=2.77, SD=0.85$ avatar, $M=1.91, SD=0.62$ robot), $t(88)=5.5547, \rho-value<0.05$. Regarding the second question, there is a significant difference between the average positive feedback scores of user experiences and the two modalities ($M=3.33, SD=0.92$ avatar, $M=4.09, SD=0.54$ robot), $t(88)=-4.72,$ $\rho-value<0.05$. Lastly, the result indicated that there is also a significant difference between the scores for which participants want to abandon the conversations with the two studies ($M=3.21, SD=1.34$ avatar, $M=1.34, SD=0.62$ robot), $t(88)=8.35,$ $\rho-value<0.05$.

\section{Discussion}

Based on these results, we observed that when participants are confused, the changes in their emotions and gaze movements after stimuli from the different conditions of the HAI study are similar to those of the HRI study, but that the changes in the range of head pose angles with different stimuli from the different conditions of the HAI study are different from those of the HRI study. Furthermore, participants prefer to engage in interaction with the robot platform rather than with the avatar in this research study and are more willing to continue to interact with the robot platform than with the avatar in this research study.

While these basic observations can be made, it is very notable that, while we attempted to unify our studies across embodiment types, it is hard in practise to achieve this. At a very technical level, our human-avatar studies were less controlled as users could participate from home -- unlike in the case of our human-robot studies. Generally, there was a very low abandonment rate for the HRI study, and we can also ensure the same quality of the dataset that we collected.

Meanwhile, it should be mentioned that during the 3-minute post-task interview, in the HAI experiment, the expectation of interaction from many participants was found to be much higher relative to the actual capabilities of the avatar. However, from the feedback of the participants in the HRI experiments, they felt fresh and curious talking to the robot, so most of them enjoyed and engaged in this HRI experiment. Also, they were surprised that the Pepper robot has high-tech social interaction skills when the Pepper robot vividly interacts with them. 

\section{Conclusion}
In this paper, we present a brief overview of a human-robot interaction study that we conducted to compliment and validate an earlier human-avatar interaction study that we conducted with equivalent tasks and feedback mechanisms. Although we attempted to ensure that experimental conditions held well across the avatar and robotic setup, in practise this was challenging to ensure. This, combined with post-interview discussions which suggested that users had relatively speaking greater disappointment with the avatar than robot, should suggest that these two platforms for interaction are not compatible when it comes to behaviour and data collection. Nevertheless, we found that there are similarities between the HAI and HRI studies based on the analysis results in the multimodal dataset that we collected: the different behaviour of the participants when they are confused or not confused is somewhat consistent in the HAI and HRI experiments in the eye-gazing estimation, emotion estimation, and self-reported confusion scores with induced confusion states. Therefore, we can see that the HAI and HRI embodiments have their own strengths and weaknesses; and that these two embodiments can potentially replace each other for the case study of confusion detection subject to high levels of control to bridge the gap between interfaces.


\begin{acks}
This publication has emanated from research conducted with the financial support of Science Foundation Ireland under Grant number 18/CRT/6183. For the purpose of Open Access, the author has applied a CC BY public copyright licence to any Author Accepted Manuscript version arising from this submission.
\end{acks}

\bibliographystyle{ACM-Reference-Format}
\bibliography{nali}

\appendix

\end{document}